\documentclass[12pt,onecolumn]{emulateapj}

\usepackage{natbib}
\usepackage{color}

\newcommand{\Area}{\mathcal A}
\newcommand{\RS}{{\rho}}
\newcommand{\uu}{{u}}
\newcommand{\Rlens}{R_\mathrm{lens}}

\newcommand{\rholens}{\rho_\mathrm{lens}}
\newcommand{\Rstar}{R_\ast}
\newcommand{\Dol}{{D_\mathrm{ol}}}
\newcommand{\Dos}{{D_\mathrm{os}}}
\newcommand{\RE}{{R_\mathrm{E}}}
\newcommand{\Afs}{{A^\ast}}
\newcommand{\Int}{\int\limits}
\newcommand{\Sum}{\sum\limits}

\shortauthors{Lee et al.}
\begin{document}
\title{Finite source effects in microlensing: A  precise, 
easy to implement, fast and numerical stable formalism }

\shorttitle{A precise, fast and stable finite-source formalism} 

\author{C.-H.~Lee\altaffilmark{1}, A.~Riffeser\altaffilmark{1}, S.~Seitz\altaffilmark{1,2} and R.~Bender\altaffilmark{1,2}}

\affil{1.University Observatory Munich, Scheinerstrasse 1, 81679 M\"unchen, Germany}
\affil{2.Max Planck Institute for Extraterrestrial Physics, Giessenbachstrasse, 85748 Garching, Germany}

\email{chlee@usm.lmu.de}

\begin{abstract}
The goal of this paper is to provide a numerically fast and stable
description  for the microlensing amplification of an extended source (either
uniform or limb-darkened) that holds in any amplification regime. We show
that our method of evaluating the amplification
can be implemented into a light-curve fitting routine using
the Levenberg-Marquardt algorithm. 
We compare the accuracy and computation times to previous methods that either
work in the high-amplification regime only, or require special treatments due to 
the singularity of elliptic integrals.

In addition, we also provide the equations including finite lens
effects in microlensing light curves.
We apply our methods to the MACHO-1995-BLG-30  and the OGLE-2003-BLG-262 events and obtain
results consistent to former studies. We derive an upper limit for the OGLE-2003-BLG-262 event lens size.

We conclude that our method allows to simultaneously search for point-source and
finite-source microlensing events in future large area microlensing surveys in a fast
manner.
\end{abstract}


\section{Introduction}

In large area microlensing surveys, one has to search for microlensing signatures in
billions of variable sources. This is straightforward to do and
computationally inexpensive in the point-source approximation.
One either fits a  Paczy{\' n}ski  light curve \citep{1986ApJ...304....1P},
or, if appropriate, the Gould high-amplification approximation for point sources \citep{1996ApJ...470..201G}.
One major disadvantage of these point-source light curves is the infinite
amplification for a lens exactly in front of the point source.

\cite{1994ApJ...421L..71G} extended Paczy{\' n}ski's light curve to finite sources
which also avoids infinite amplifications. His equation describes the amplification
as the two-dimensional integration of the Paczy{\' n}ski  amplification over the
circular source, assumed to have constant surface brightness.

Using the limiting form of Paczynski light curve under high amplification, 
Gould is able to factor out the two-dimensional integral into point source amplification 
times a much simpler integral. Meanwhile, \cite{1994ApJ...430..505W} obtained the finite 
source amplification directly from the lens equation
by comparing the area of the source and its lensed images. However, one needs to take care 
of the singular points for the elliptic integrals of the first and the third kind when 
using their formula.

In this paper we adopt the same strategy as \cite{1994ApJ...421L..71G}, 
because in this way more general surface-brightness profiles for the sources 
(e.g. limb-darkened ones) can be taken into account straightforwardly. For the cases 
of a uniform disk we will also compare our results with \cite{1994ApJ...430..505W}.

The technical issue of the integration in the Gould extended source
formalism can be carried out in several different ways.
The two straightforward ones are to use polar coordinates
and to choose  the coordinate center either (1) at the source center or (2)
at the lens center. \cite{1994ApJ...421L..71G} took the first choice.
\cite{2006ApJS..163..225R} have shown for the very special
case where the lens is positioned along the line of sight to the source
that the integration can be solved very easily if the second option is chosen.

This leads us to choose the lens center as the coordinate center in general
to benefit from the more simple integrand. We will show (in Section 2)
that in this way the amplification of a uniform circular source is
reduced to a one-dimensional integral and can be computed numerically fast and stable
by using the composite Simpson's rule.
A limb-darkened source is treated in Section 3. The two-dimensional
integral can be solved numerically again in a fast and stable fashion,
and light-curve fits for limb-darekened profiles can be obtained with
the Levenberg-Marquardt algorithm \citep[see][]{2007nrc..book.....P} with
less than 100 steps.
We also allow for finite lens sizes in Section 4.
As a test example, we apply our fitting methods to a MACHO \citep{1992ASPC...34..193A} 
event and an OGLE \citep[The Optical Gravitational Lensing Experiment;][]{1992AcA....42..253U} event in
Section 5. We conclude in Section 6.

\begin{figure}[t]
  \centering
  \epsscale{1.}
  \plottwo{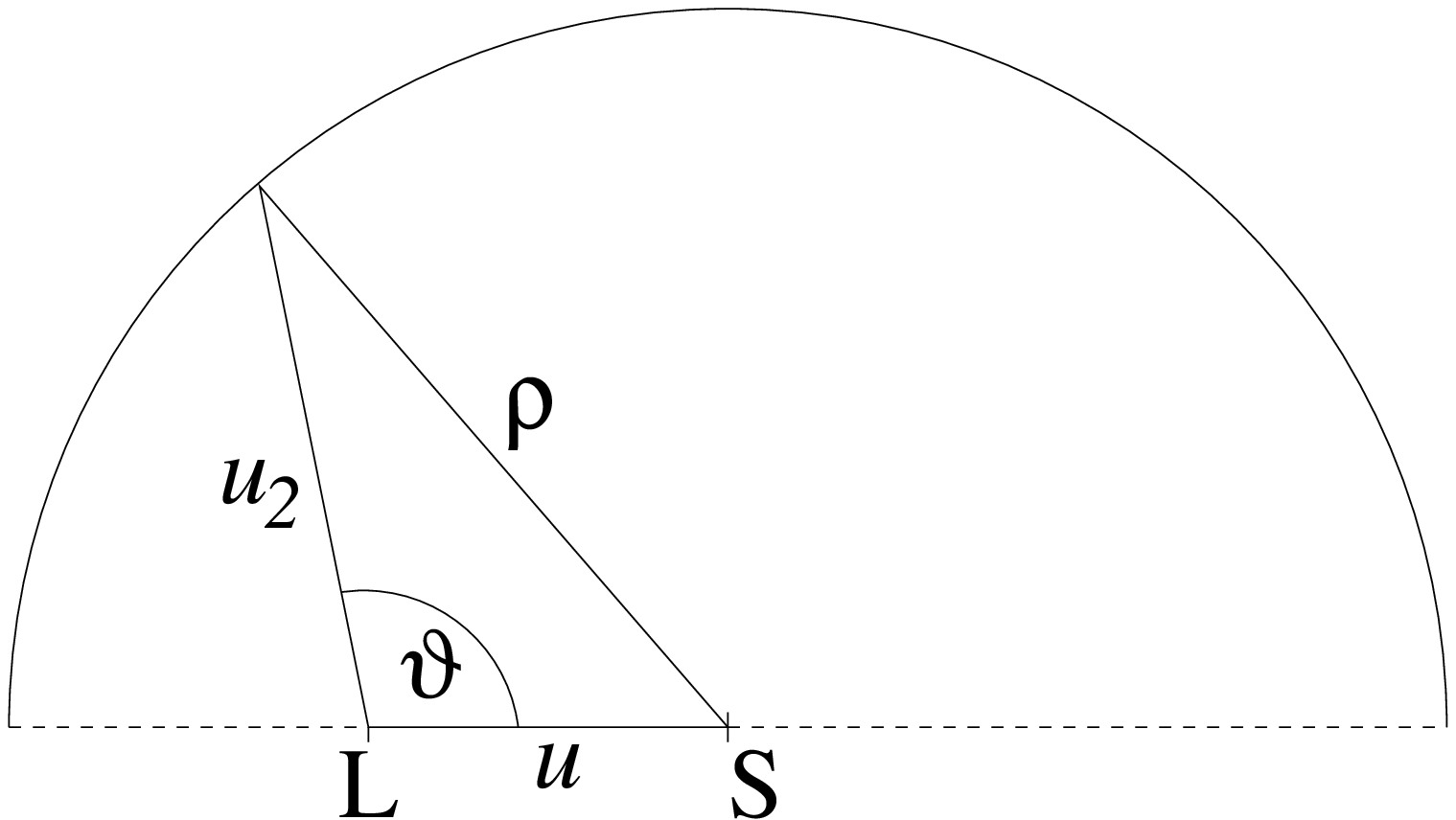}{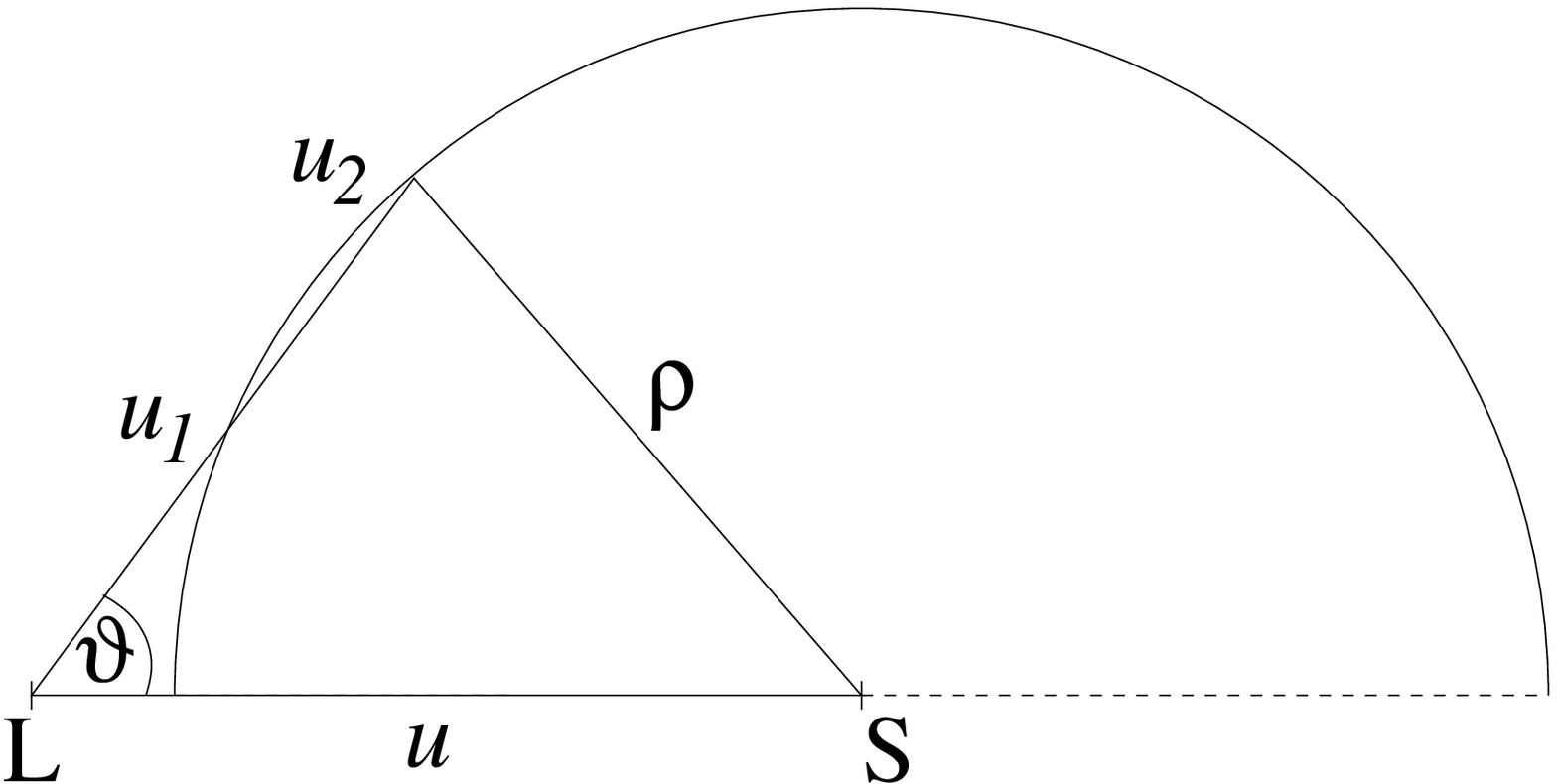}
  \caption{Geometric definitions. Left: source is overlapping the lens center. Right: lens is outside the source
    radius.}
  \label{fig.fs_geometry}
\end{figure}

\section{The finite-source microlensing equation}

We first introduce our notation. Let $R_E$ be the Einstein radius of a point mass lens, and 
$b$ be the impact parameter of a point source. Then one can write the amplification of the 
point source by the point mass lens as a function of the dimensionless impact parameter $u:=b/R_E \equiv \theta/\theta_{E}$ as
\begin{equation}
A_{_{\mathrm{PS}}}(u) = \frac{u^2+2}{u\sqrt{u^2+4}}
 \label{eq.A_pac}
\end{equation}
\citep{1986ApJ...304....1P}.

If the source is extended one can obtain the lensed flux and the total amplification by integrating 
$A(u)$ over the source area, weighted by the surface-brightness profile of the source.
We now derive the amplification for a circular source\footnote{The reader is referred to \cite{1997ApJ...490...38H} 
for a more general case of elliptical source.} with radius $\RS$ ($\RS\equiv\frac{\Rstar\Dol}{\RE\Dos}$ 
is the projected source size in units of the Einstein radius $R_E$ and $\Rstar$ is the physical source size). 
The situation is sketched in Figure~\ref{fig.fs_geometry}. There are two cases: either the center of the 
lens $L$ (projected along the line of sight) is within the extended source centered at $S$ (the left side of 
Figure~\ref{fig.fs_geometry}) or the lens is outside the extended source (the right side of Figure~\ref{fig.fs_geometry}), 
i.e., either $u\le\RS$ or $u>\RS$.
One obtains the amplification of the extended uniform source by integrating the point-source amplification 
over the source area $\Area_{\mathrm{source}}$: 
\begin{equation}
\Afs(u;\RS)=\frac{\Int_{\Area_{\mathrm{source}}} A_{_{\mathrm{PS}}}d\Area}{\Int_{\Area_{\mathrm{source}}}d\Area} = \frac{1}{\pi\RS^2}\Int_{\Area_{\mathrm{source}}}A_{_{\mathrm{PS}}}d\Area~.
\end{equation}

Using polar coordinates centered on the lens, one can write
\begin{equation}
  \begin{array}{l}
    \Afs(u;\RS)
    = \frac{2}{\pi\RS^2} 
    \Int_0^{\pi} \,
    \Int_{u_1(\vartheta)}^{u_2(\vartheta)} 
    A_{_{\mathrm{PS}}}(\tilde{u})\, \tilde{u}\,d\tilde{u} \,d\vartheta ~.
  \end{array}
  \label{eq.A_fin}
\end{equation}

The integration boundaries $u_1$ and $u_2$ are
\begin{equation}
  u_1(\vartheta) =
  \left\{
  \begin{array}{ll}
    0
    &
    ,u\le\RS
    \\
    u \cos{\vartheta} - \sqrt{\RS^2 - u^2 \sin^2\vartheta}
    &
    ,u>\RS \wedge \vartheta\le\arcsin(\RS/u)
    \\
    0
    &
    ,u>\RS \wedge \vartheta > \arcsin(\RS/u)
  \end{array}
  \right.
~,
  \label{eq.u1}
\end{equation}

\begin{equation}
  u_2(\vartheta) =
  \left\{
  \begin{array}{ll}
    u \cos{\vartheta} + \sqrt{\RS^2 - u^2 \sin^2\vartheta}
    &
    ,u\le\RS
    \\
    u \cos{\vartheta} + \sqrt{\RS^2 - u^2 \sin^2\vartheta}
    &
    ,u>\RS \wedge \vartheta\le\arcsin(\RS/u)
    \\
    0
    &
    ,u>\RS \wedge \vartheta > \arcsin(\RS/u)
  \end{array}
  \right.
~,
  \label{eq.u2}
\end{equation}
and so the amplification becomes
\begin{equation}
\Afs(u;\RS)= \frac{1}{\pi\RS^2} 
\Int_0^{\pi} \left[u_2(\vartheta)\sqrt{u_2(\vartheta)^2+4} - u_1(\vartheta)\sqrt{u_1(\vartheta)^2+4} \right]\,d\vartheta
\label{eq.Afs}
~,
\end{equation}
which can be approximated numerically using the composite Simpson's rule with $n$(an even number) grids:
\begin{equation}
  \Afs(u;\RS)\approx
  \left\{
    \begin{array}{ll}
      \scriptstyle{\frac{1}{\pi\RS^2}\frac{\pi}{2n}\left[\frac{\left(u+\RS\right)\sqrt{\left(u+\RS\right)^2+4}-\left(u-\RS\right)\sqrt{\left(u-\RS\right)^2+4}}{3} + \frac{2}{3}\Sum_{k=1}^{n-1}f\left(\frac{2k\pi}{2n}\right) + \frac{4}{3}\Sum_{k=1}^{n}f\left(\frac{(2k-1)\pi}{2n}\right) \right]} &,u\le\RS\\
\scriptstyle{\frac{1}{\pi\RS^2}\frac{\arcsin(\RS/u)}{n}\left[\frac{\left(u+\RS\right)\sqrt{\left(u+\RS\right)^2+4}-\left(u-\RS\right)\sqrt{\left(u-\RS\right)^2+4}}{3}+\frac{2}{3}\Sum_{k=1}^{n/2-1}f\left(\frac{2k\arcsin(\RS/u)}{n}\right) +\frac{4}{3}\Sum_{k=1}^{n/2}f\left(\frac{(2k-1)\arcsin(\RS/u)}{n}\right) \right]} &,u>\RS\\
    \end{array}
  \right.
  ~,
  \label{eq.A_discret}
\end{equation}

where $f(\vartheta) = [u_2(\vartheta)\sqrt{u_2(\vartheta)^2+4} - u_1(\vartheta)\sqrt{u_1(\vartheta)^2+4}]$. The upper limit of $\vartheta$ 
changes from $\frac{\pi}{2}$ to $\pi$ when the lens crosses the edge of the source from outside to inside, thus we set a grid of $2n$ for $u\le\RS$ 
in order to have the same step size on both sides.   

\begin{figure}[th]
  \centering
\epsscale{1.1}
\plottwo{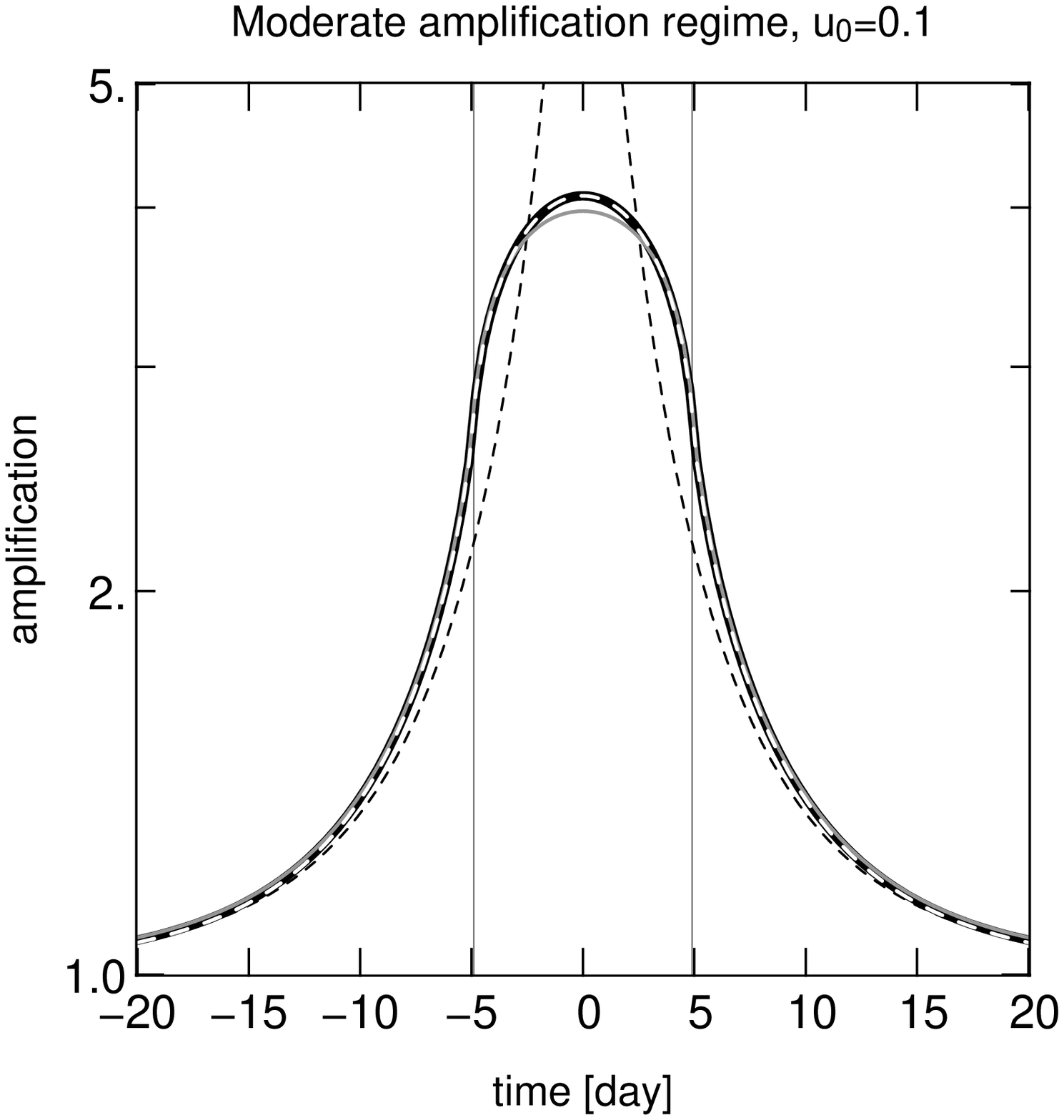}{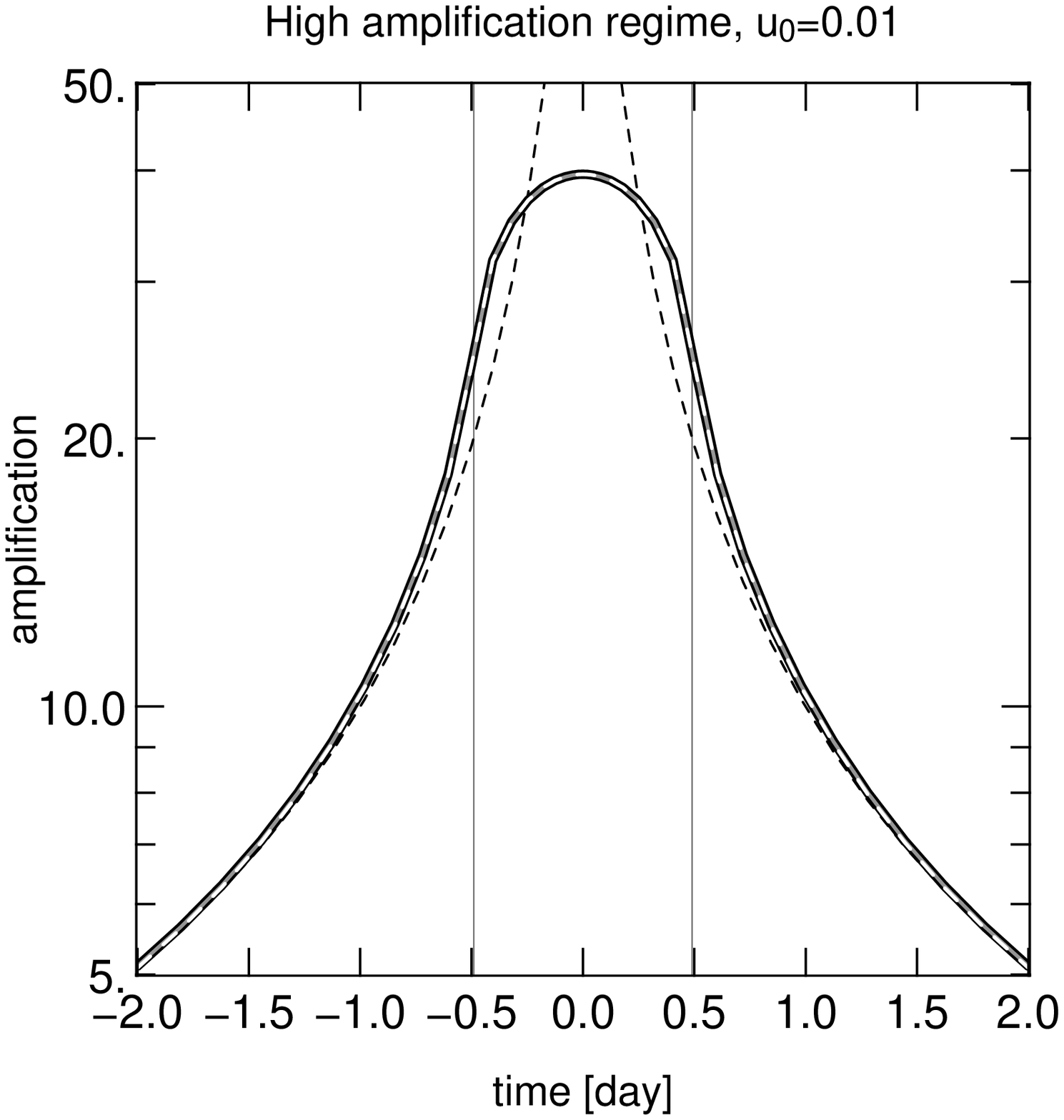}
  \caption{Comparison of finite-source light-curve approximations. Left: moderate-amplification regime with $t_E = 10$, $u_0 = 0.1$ and $\RS = 0.5$. 
Right: high-amplification regime with $t_E = 10$, $u_0 = 0.01$ and $\RS = 0.05$. 
    In \textit{dashed black} the Paczy{\' n}ski light curve for a point source, in \textit{solid black} 
Witt \& Mao light curve, 
in \textit{gray} the approximation derived by \cite{1994ApJ...421L..71G} and in
    \textit{dashed white} Equation~(\ref{eq.A_discret}) with $n=10$. The vertical 
lines indicate the time when $u = \RS$. Our formula is as good as \cite{1994ApJ...421L..71G} in high-amplification 
regime and is better in the moderate-amplification regime. }
  \label{fig.Afs_example}
\end{figure}

\begin{figure}[h]
  \centering
  \epsscale{0.7}
  \plotone{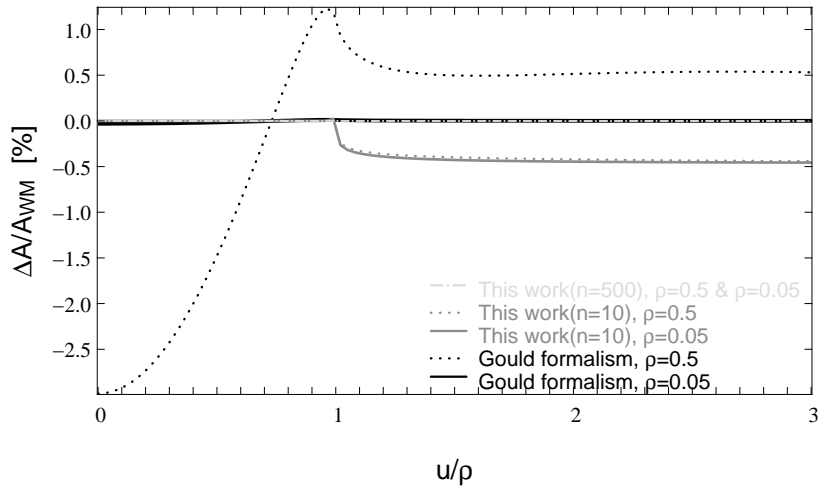}
\caption{Percentage deviation in amplification compared to Witt \& Mao formalism ($A_{_{\mathrm{WM}}}$). 
The expression of \cite{1994ApJ...421L..71G} is valid for small source (\textit{solid black}) 
but shows deviation $>$ 2.5\% for larger source (\textit{dotted black}). Equation~(\ref{eq.A_discret}) 
with $n$ = 10 shows a smaller deviation ($<$ 0.5\%). Equation~(\ref{eq.A_discret}) with $n$ = 500 for both source 
sizes are well overlapped with each other, so we show here only $\rho = 0.05$.}
  \label{fig.A_fs_zoomin}
\end{figure}

\cite{1994ApJ...421L..71G} argued that the finite-source effects are prominent only when the lens 
is very close to the source center ($u \ll 1$), and thus one can approximate Equation~(\ref{eq.A_pac}) by

\begin{equation}
  \begin{array}{ll}
    A_{_{\mathrm{PS}}}(u) = \frac{u^2+2}{u\sqrt{u^2+4}}\approx u^{-1} & ,u\ll1~,
  \end{array}
  \label{eq.A_gould}
\end{equation} 

and the finite-source light curve can be obtained by solving elliptic integrals \citep[see also][]{2004ApJ...603..139Y,2006A&A...460..277C}

\begin{equation}
  \Afs_{_{\mathrm{Gould}}}(u;\RS) \simeq  
  A_{_{\mathrm{PS}}}(u)\frac{4u}{\pi\RS}E\left(\vartheta_{\max},\frac{u}{\RS}\right)~,
  \label{eq.A_gould_elliptic_integrals}
\end{equation}

where $E(\phi,k)$ is the elliptic integral of the second kind and $\vartheta_{\max}$ is defined as
\begin{equation}
  \vartheta_{\max} =
  \left\{
    \begin{array}{ll}
      \frac{\pi}{2} &, u \le \RS
      \\
      \arcsin(\RS/u) &, u > \RS
    \end{array}
  \right.\\
  .
  \label{eq.Theta_max}
\end{equation}
\\
We now compare our method for $\Afs(u;\RS)$ with previous ones,
i.e. with \cite{1986ApJ...304....1P}, \cite{1994ApJ...421L..71G} and \cite{1994ApJ...430..505W},
and illustrate these comparisons in Figures~\ref{fig.Afs_example} and
\ref{fig.A_fs_zoomin}.  
\\ 
Equation~(\ref{eq.A_gould_elliptic_integrals}) and (\ref{eq.Theta_max}) 
allow a fast computation of finite-source light curves in the Gould approximation, which however is
accurate only for a high-amplification event. This is shown in
Figure~\ref{fig.Afs_example}, where for high amplifications (right
panel) the Gould finite-source approximation (\textit{gray}) is very close to
the \cite{1994ApJ...430..505W} light curve (displayed in \textit{solid black}), but fairly off when the lens transits the source for
low amplifications (left panel).

Our formalism from Equation~(\ref{eq.Afs}) and that of \cite{1994ApJ...430..505W} both
provide the exact light curves for uniform extended sources. In the
Witt \& Mao formalism one has to evaluate an elliptic integral
which shows singularity when the impact parameter $u$ is similar 
to the source size $\rho$. Witt \& Mao therefore derived a separate 
solution for the case of $u=\rho$. This method is
difficult to implement into numerical fitting routines in general, 
and particular cumbersome for those fast numerical fitting routines, 
where the partial derivatives have to be provided.

We therefore suggest to start from our exact formalism given in
Equation~(\ref{eq.Afs}) and estimate values for the integral using
Equation~(\ref{eq.A_discret}) with $n=10$. The comparison with results from
higher values for $n=500$ or the comparison with the \cite{1994ApJ...430..505W} 
formalism -- see the \textit{gray} and \textit{dash-dotted curves} in Figure~\ref{fig.A_fs_zoomin} -- shows
that Equation~(\ref{eq.A_discret}) (with $n=10$) provides a precise numerical
estimate for the integral already. Another advantage of our formalism
is that one can obtain the derivatives of Equation~(\ref{eq.Afs})
with respect to source radius $\RS$ and $\uu$ in a straightforward
manner (see Appendix A).  This enables us to use fitting routines as,
e.g., the Levenberg-Marquardt algorithm \citep[see][]{2007nrc..book.....P}
which converge in this case in less than 100 iterations.

\begin{figure}[!h]
\begin{minipage}[b]{0.48\linewidth}
\centering
\epsscale{1.2}
\plotone{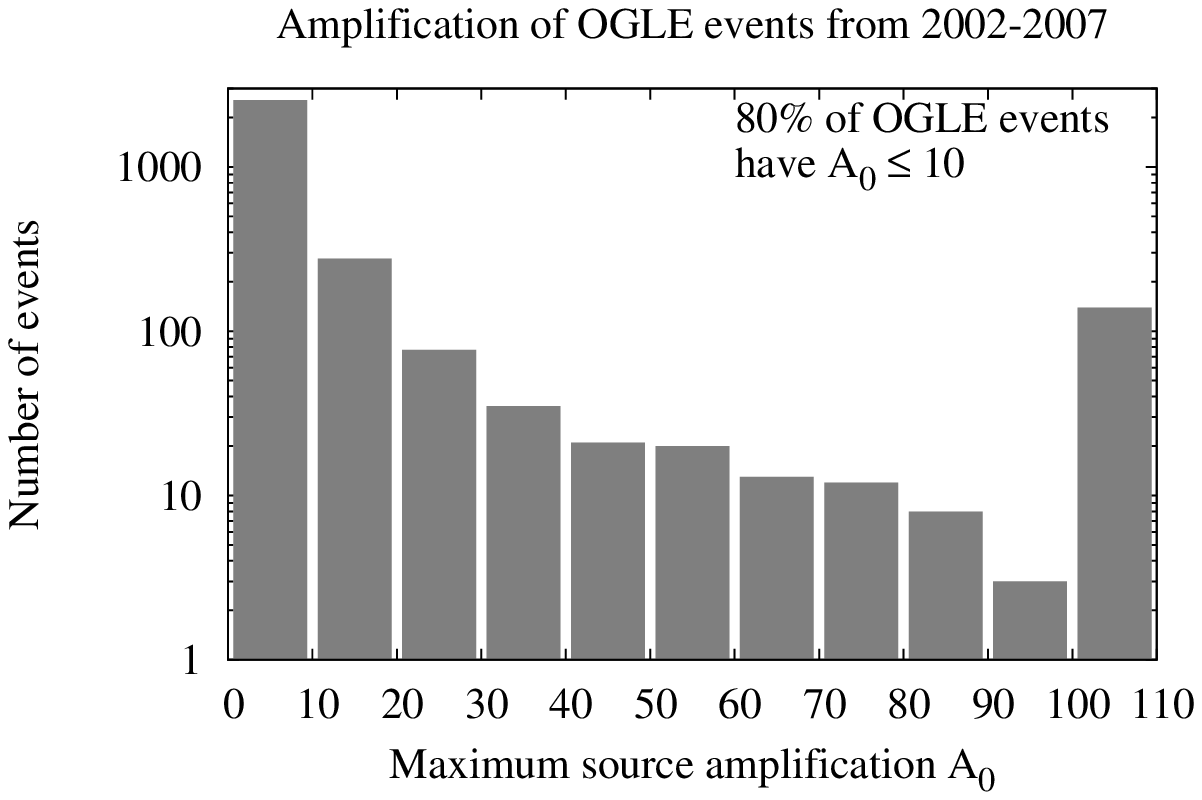}
\caption{Maximum amplification of microlensing events detected by the OGLE experiment from 2002 to 2007. Most of the events ($>80\%$) have 
    maximum amplification $<$ 10. Events with maximum amplification $>$ 100 , which are categorized into interval 100-110 in this plot, are relatively rare ($<4.4\%$).}
\label{fig.OGLE}
\end{minipage}
\hspace{0.5cm}
\begin{minipage}[b]{0.48\linewidth}
\centering
\epsscale{1.2}
\plotone{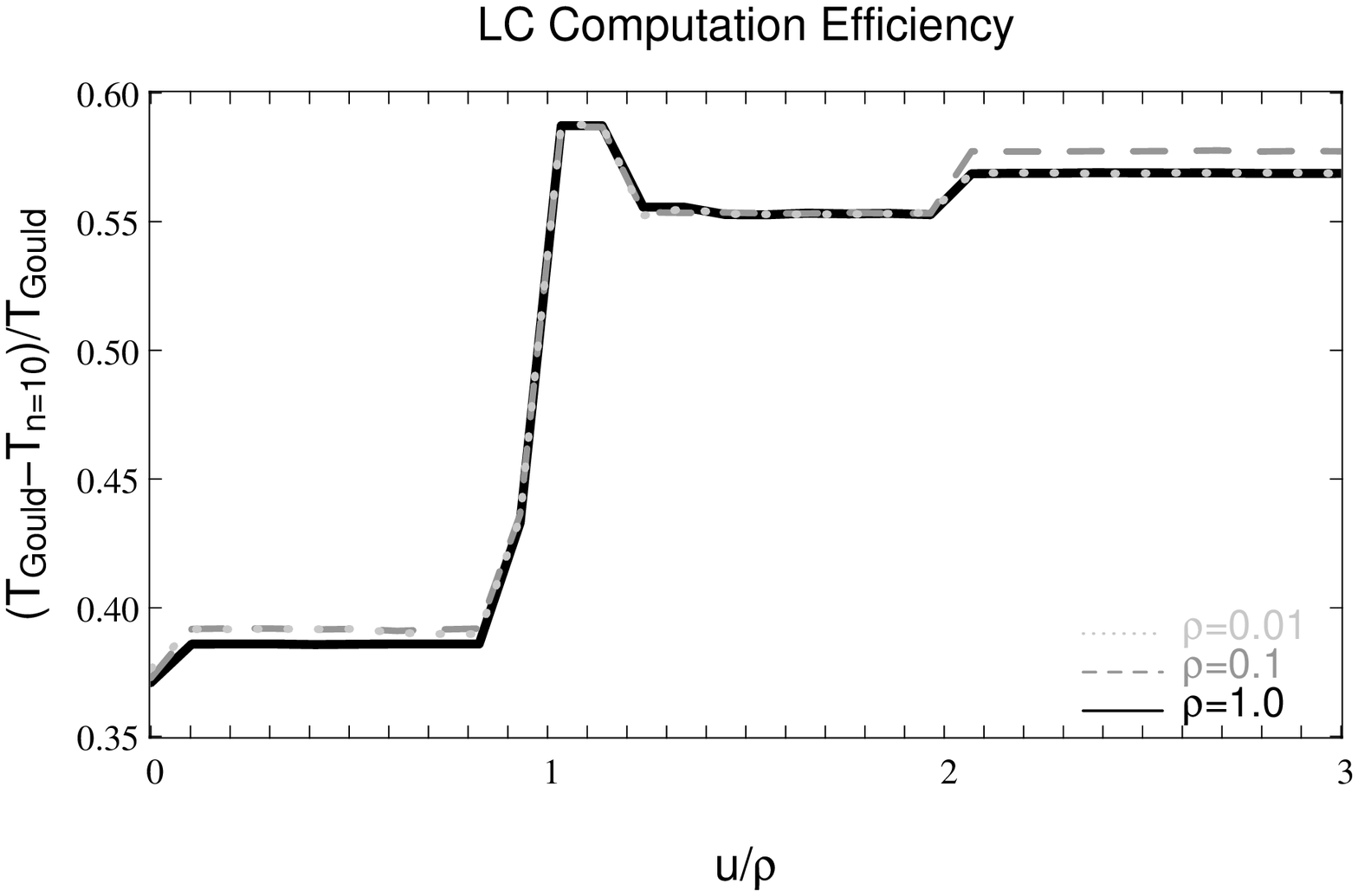}
\caption{Light-curve computation efficiency. We compare light-curve computation time of Equation~(\ref{eq.A_discret}) with $n = 10$ 
    to that of Gould's formalism for various source radii ($\RS$ = 0.01, 0.1, and 1). The computation time for our approximation is comparable to the
    Gould formalism; it is about $38\%$ faster when $u<\RS$ and is $>55\%$ faster when $u>\RS$.}
\label{fig.CT}
\end{minipage}
\end{figure}

The approximation by \cite{1994ApJ...421L..71G} with $A_{_{\mathrm{PS}}}(u)$ evaluated according to 
Equation~(\ref{eq.A_pac}) is actually valid for all $u$ provided that $\rho\ll1$, so it deviates from 
Equation~(\ref{eq.Afs}) for larger source size. 
In fact, more than 80\% (2548 out of 3153) of the microlensing events detected from the 
OGLE experiment\footnote{http://ogle.astrouw.edu.pl/ogle3/ews/ews.html} \citep{2003AcA....53..291U} 
have maximum amplification $<$ 10 (see Figure~\ref{fig.OGLE}). This highlights the 
necessity of a fast fitting routine for the moderate-amplification regime. We then compare the 
light-curve computation time of Equation~(\ref{eq.A_discret}) to Gould's formalism 
(see Figure~\ref{fig.CT}). With $n=10$, Equation~(\ref{eq.A_discret}) is about 38$\%$ faster then Gould's 
formalism when $u\le\RS$ and is $>$ 55$\%$ faster when $u>\RS$. Therefore, our $n=10$ approximation 
turns out to be a practical fast fitting routine for both moderate- and high-amplification regimes.

\section{Finite source with limb darkening}
The next step towards a more precise microlensing light curve for extended sources is to account
for limb darkening. Since the darkening is increasing towards the edges of the source, the limb
darkening brings finite-source light curves closer to the Paczy{\' n}ski
light curve which can be considered as the most extreme limb-darkening
model with a delta function.

We use the one-parameter linear limb-darkening profile from \cite{2004ApJ...603..139Y} 
for the surface brightness of the source,
\begin{eqnarray}
\label{eq-limbd1}
S(r/R_*,\Gamma) = \bar{S}\left[1-\Gamma\left(1-\frac{3}{2}\sqrt{1-\left(\frac{r}{R_*}\right)^2}\right)\right] ~,
\end{eqnarray}

where $r$ is the distance to the source center. 

$\Gamma$ is the 
limb-darkening coefficient, and depends on the wavelength range used for the observations. 
$\bar{S}$ is the mean surface brightness of the source and defined as
\begin{equation}
\Int_0^{2\pi}\Int_0^{R_*}S(\tilde{r}/R_*,\Gamma)\tilde{r}d\tilde{r}d\vartheta=\pi{R_*}^2\bar{S}~.
\end{equation}

We implemented the limb-darkening effects in our finite-source light curve as
follows:

\begin{equation}
  \begin{array}{rl}
    \Afs_{_{\mathrm{LD}}}(u;\RS)
    &= \frac{2}{\pi\RS^2\bar{S}} 
    \Int_0^{\pi} \,
    \Int_{u_1(\vartheta)}^{u_2(\vartheta)} 
    A_{_{\mathrm{PS}}}(\tilde{u})\,S(r/R_*,\Gamma)\,\tilde{u}d\tilde{u} \,d\vartheta
\\
    &= \frac{2}{\pi\RS^2} 
    \Int_0^{\pi} \,
    \Int_{u_1(\vartheta)}^{u_2(\vartheta)} 
    \frac{\tilde{u}^2+2}{\sqrt{\tilde{u}^2+4}}\,
    \left[
      1-\Gamma
      \left(
      1-\frac{3}{2}{\sqrt{1-\frac{\tilde{u}^2-2u\tilde{u}\cos\vartheta+u^2}{\RS^2}}}
      \right) 
      \right]\,
    d\tilde{u} \,d\vartheta
~.
  \end{array}
  \label{eq.A_fin_ld}
\end{equation}

\begin{figure}[h]
  \centering
\epsscale{0.8}
\plotone{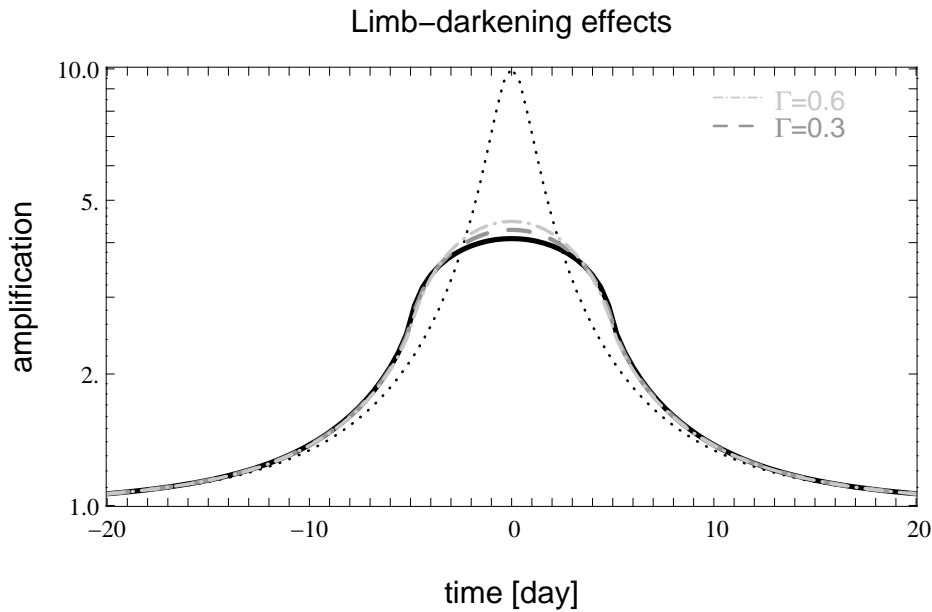}
  \caption{Limb-darkening effects on the finite-source light curve in the  
moderate-amplification regime. In \textit{dotted black} we show the Paczy{\' n}ski light 
curve for a point source with $t_E$ = 10 and $u_0$ = 0.1. In 
\textit{solid black}, we show the finite-source light curve for a uniform source with a projected source size of $\RS$ = 0.5. In \textit{dashed line} and \textit{dash-dotted line}, we plot the limb-darkened 
finite-source light curves with $\Gamma$ = 0.3 and 0.6. Increasing $\Gamma$ 
enhances the limb-darkening effects thus brings the finite-source light curve 
closer to Paczy{\' n}sky's formalism.} 
  \label{fig.LD}
\end{figure}

Equation~(\ref{eq.A_fin_ld}) is still a double integral over $\tilde{u}$ and 
$\vartheta$. But even here the divergent part cancels, and the function is numerically stable and can be 
evaluated using a small grid. The limb-darkening effects under moderate-amplification regime is shown in Figure~\ref{fig.LD}.

\begin{figure}[h]
  \centering
\epsscale{0.7}
\plotone{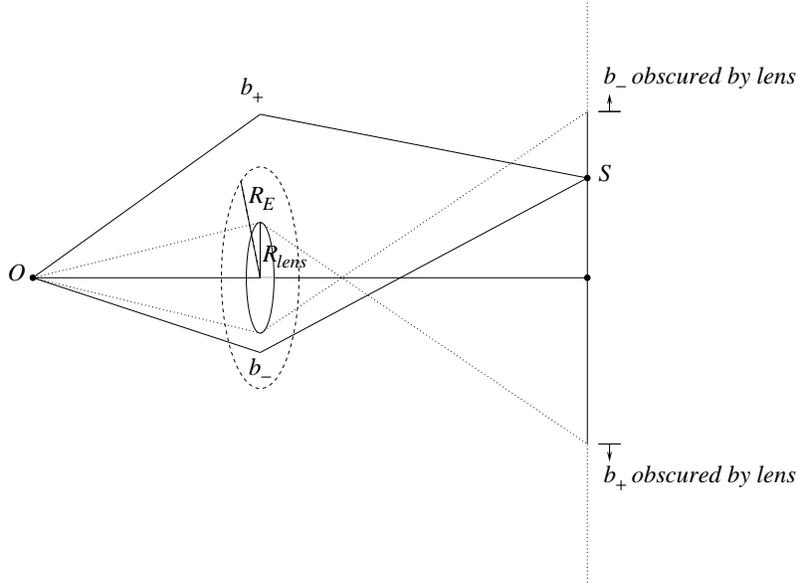}
  \caption{Image obscuration by a  finite lens with radius $R_{lens} = 0.5 R_E$.} 
  \label{fig.lens05RE}
\end{figure}

\section{Finite-source equation with finite lens}
Given a finite-size lens, one can always find a time interval when the lens obscures the inner 
(and the outer, depending on the lens size) lensed image in the early rising stage and in the 
final declining stage of the light curve. In the following, we investigate how large this effect 
is depending on the lens size.

\cite{2002ApJ...579..430A} derived the lens-modified amplification by calculating how much area 
is unobscured by the lens in the image plane. One has to solve for the image position by inverting 
the lens equation and one has to evaluate the image area from the image boundary using Stokes' theorem. 
Depending on the source and lens radii, there are 7 different cases for the inner image and 6 cases 
for the outer image to be considered if one follows the derivation of Agol. 
    
Here we show that the finite lens amplification of a finite source again can be 
much more easily evaluated if one uses the polar coordinates $\tilde{u}$ and $\vartheta$ again.
First, we consider a lens with physical radius ${R_\mathrm{lens}}$ transiting the 
surface of the source. The light emitted at a given point from the source follows the lens equation  
\begin{equation}
      \frac{b}{R_E} = \frac{b_{\pm}}{R_E} - \frac{R_E}{b_{\pm}} \\ 
\label{eq.lenseq}
~,
\end{equation}
which gives the position of the two images in the lens plane (recall $u:=\frac{b}{R_E}$) 
\begin{equation}
\frac{b_{\pm}}{R_E}=\frac{u\pm\sqrt{u^2+4}}{2} \\
\label{eq.sol_lenseq}
\end{equation}
with amplifications
\begin{equation}
A_{\pm}(u)=\frac{u^2+2}{2u\sqrt{u^2+4}}\pm\frac{1}{2}\\
\label{eq.Apm}
~.
\end{equation}

Here, $\frac{b_{+}}{R_E}$ denotes the outer image, and $\frac{b_{-}}{R_E}$ denotes the inner image in units 
of the Einstein radius. The sum of $A_{+}(u)$ and $A_{-}(u)$ gives the Paczy{\' n}ski light curve.

An image is unobscured if $-\frac{b_{-}}{R_E} > \rholens$ or $\frac{b_{+}}{R_E} > \rholens$ holds, 
where $\rholens \equiv \frac{\Rlens}{\RE}$ is the lens radius in units of the Einstein radius.
Following this criterium and Figure~\ref{fig.lens05RE}, there exists an upper limit 
for $\frac{b_{-}}{R_E} < -\rholens$ and a lower limit for $\frac{b_{+}}{R_E} > \rholens$ 
to be unobscured by the lens. Therefore, we only need to consider these two limitations when 
integrating the amplification in Equation~(\ref{eq.A_fin}): 

\begin{equation}
  \begin{array}{rl}
    \Afs_{_{\mathrm{FL}}}(u;\RS)
    &
    = \frac{2}{\pi\RS^2} 
    \Int_0^{\pi} \,
    \Int_{u_1(\vartheta)}^{u_2(\vartheta)} 
    \left[
      A_+(\tilde{u})\Theta\left(\frac{b_{+}}{R_E}-\rholens\right)+
      A_-(\tilde{u})\Theta\left(-\frac{b_{-}}{R_E}-\rholens\right)
      \right]
    \tilde{u}\,d\tilde{u} \,d\vartheta
\\
&
  = \frac{1}{\pi\RS^2} 
  \Int_0^{\pi} \, 
  \left[
    \left.
      \left(
        \frac{\tilde{u}}{2}\sqrt{\tilde{u}^2+4} + \frac{\tilde{u}^2}{2}
      \right) 
    \right|_{\max\left[u_1(\vartheta),\rholens-\frac{1}{\rholens}\right]}^{u_2(\vartheta)}+
    \left.
      \left(
        \frac{\tilde{u}}{2}\sqrt{\tilde{u}^2+4} - \frac{\tilde{u}^2}{2} 
      \right)
    \right|_{u_1(\vartheta)}^{\min\left[u_2(\vartheta),-\rholens+\frac{1}{\rholens}\right]} 
  \right]
  \,d\vartheta~,
\end{array}
  \label{eq.finite_lens}
\end{equation}

where $\Theta(x)$ defines the Heaviside step function.

Combining Equation~(\ref{eq.A_fin_ld}) and Equation~(\ref{eq.finite_lens}) fully considers a limb-darkened source and a finite lens: 
\begin{equation}
\Afs_{_{\mathrm{LD\&FL}}}(u;\RS)
= \frac{2}{\pi\RS^2} 
\Int_0^{\pi} \,
\Int_{u_1(\vartheta)}^{u_2(\vartheta)} 
\left[
  A_+(\tilde{u})\Theta\left(\frac{b_{+}}{R_E}-\rholens\right)+
  A_-(\tilde{u})\Theta\left(-\frac{b_{-}}{R_E}-\rholens\right)
  \right]
S(r/R_*,\Gamma)
\tilde{u}\,d\tilde{u} \,d\vartheta
~.
\end{equation}

\section{Results}
When we implemented the finite-source fitting using the Levenberg-Marquardt
algorithm, we recognized that a good set of initial values is
needed to bring the algorithm to convergence. Fitting a Paczy{\' n}ski light
curve to derive these initial values for the finite-source fitting
leads to very good results. The algorithm is stable and for an initial
value of $\RS=0.1$ it converges within 100 iterations.

\cite{1997ApJ...491..436A} were able to measure a microlensing light
curve with finite-source effects in MACHO-1995-BLG-30. We extracted the data points from their paper and applied our 
finite-source fitting algorithms to them. Fitting Equation~(\ref{eq.A_discret}) with $n=10$ to the data yields a 
perfect agreement (see Table~\ref{tab.fit} and Figure~\ref{fig.MACHO9530}) with the parameters given in Table 2 
of \cite{1997ApJ...491..436A}: 

\begin{figure}[!th]
  \centering
\epsscale{0.74}
\plotone{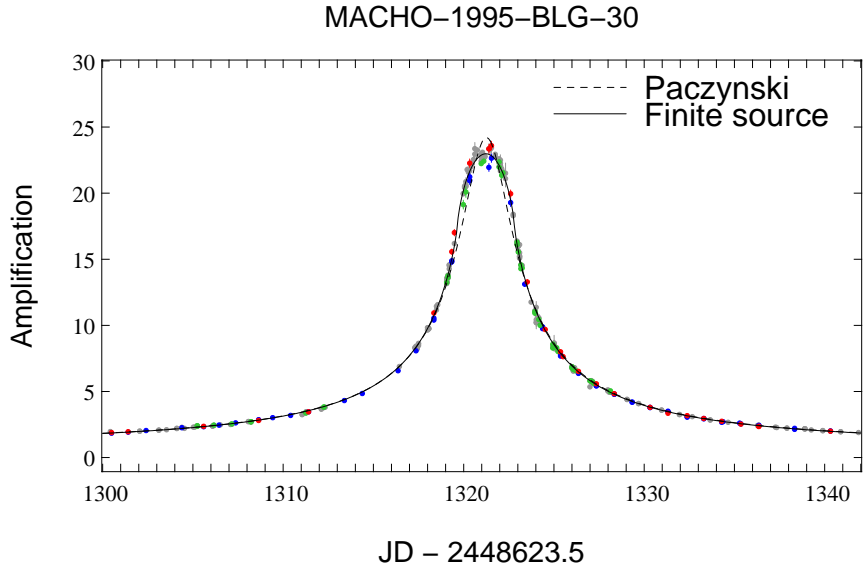}
  \caption{Finite-source light-curve fits for MACHO-1995-BLG-30 assuming a uniform source. Data points in \textit{R} are from MACHO (\textit{red}), 
    CTIO, UTSO, WISE, and MJUO (\textit{gray}) and \textit{V} are from MACHO (\textit{blue}) and UTSO (\textit{green}). 
    The \textit{dashed line} shows the light curve for a point-source model. The best-fitting finite-source light-curve parameters 
are displayed in Table~\ref{tab.fit}   
}
  \label{fig.MACHO9530}
\end{figure}

\begin{table}[!h]
\caption{Light-curve parameters for MACHO-1995-BLG-30}
\footnotesize
\center
\begin{tabular}[t]{llll}
\hline
Fit      & $A_{_{\mathrm{PS}}}(u)$~(this work) & $\Afs(u;\RS)$~(this work) &  \cite{1997ApJ...491..436A}, Table 2\tablenotemark{a}\\ 
\hline
t$_0$    & 1321.260 $\pm$ 0.002     & 1321.235 $\pm$ 0.002     & 1321.2(1)   \\
t$_E$    & 34.41 $\pm$ 0.02         & 34.25 $\pm$ 0.02         & 33.68(1)    \\
u$_0$    & 0.04133 $\pm$ 0.00004    & 0.05569 $\pm$ 0.00006    & 0.05579(1)  \\
$\RS$    & --                       & 0.0722 $\pm$ 0.0001      & 0.07335(1)  \\
\hline

\end{tabular}
\tablenotetext{a}{The reported uncertainties in the final significant digit(s) of \cite{1997ApJ...491..436A} are the maximum extent of the surface in parameter space which has a $\chi^2$ greater than the best-fit value by 1.}
\label{tab.fit}
\end{table}
\normalsize

\begin{figure}[!h]
  \centering
\epsscale{0.74}
\plotone{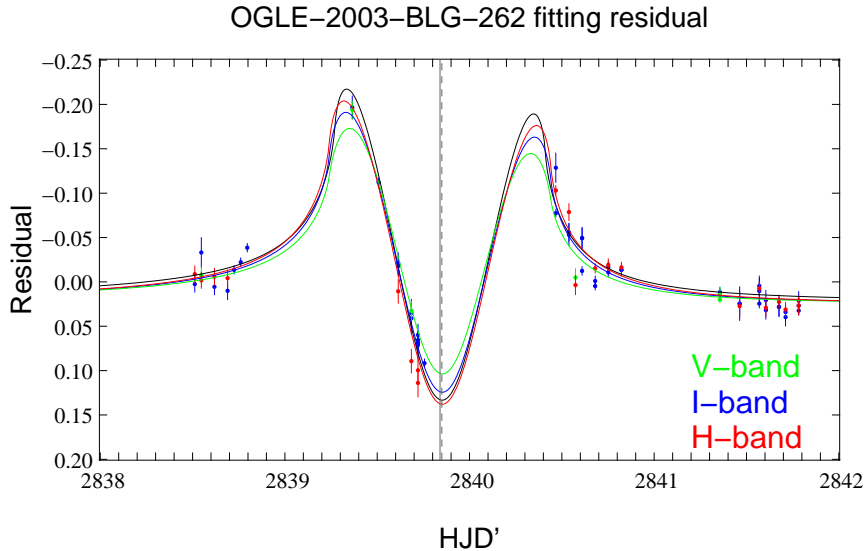}
  \caption{Residuals of the observed light curve relative to the best-fitted point-source light curve. The \textit{solid black} curve shows the light curve of an extended
    source with uniform surface brightness. The \textit{solid blue}, \textit{solid red}, and \textit{solid green} curves are 
    extended source models incorporating limb darkening in $V$, $I$ and $H$ bands with $(\Gamma_V, \Gamma_I, \Gamma_H) = (0.72, 0.44, 0.26)$. 
    The vertical lines indicate $t_0$ for the best-fitted point-source (\textit{solid}) and limb-darkened finite-source (\textit{dashed}) model. 
    For the light curves with the limb-darkened source we have left $t_0$ as a free parameter. The best-fitting value for $t_0$ slightly 
    differs (see Table~\ref{tab.fit_OGLE262}). This causes the asymmetric pattern of the residual relative to the Paczy{\'n}ski light curve.
}
  \label{fig.OGLE262_residual}
\end{figure}

\begin{figure}[h]
\begin{minipage}[b]{0.5\linewidth}
\centering
\epsscale{1.15}
\plotone{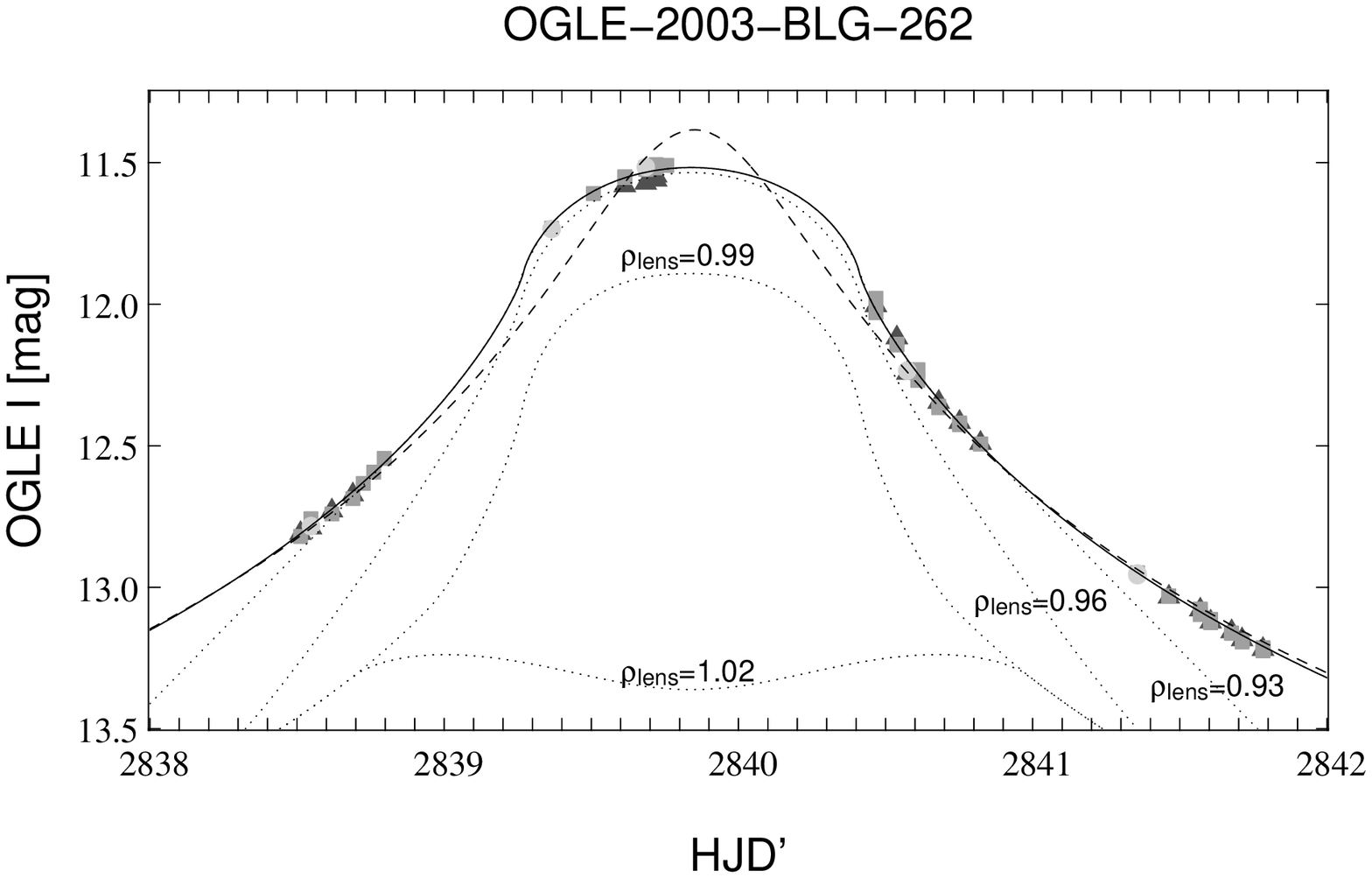}
  \caption{Finite-source and finite-lens light-curve fits for OGLE-2003-BLG-262. Data points are in \textit{I}(square), \textit{V}(circle), and \textit{H}(triangle). 
    The \textit{dashed line} shows the light curve for a point source. The \textit{solid line} shows the light curve for an extended source with uniform surface brightness. 
    The \textit{dotted lines} illustrate the effects of finite lens sizes on top of finite-source size for lens sizes of $\rholens$ = 0.93, 0.96, 0.99, and 1.02.
}
  \label{fig.OGLE262}
\end{minipage}
\hspace{0.5cm}
\begin{minipage}[b]{0.5\linewidth}
\centering
\epsscale{1.15}
\plotone{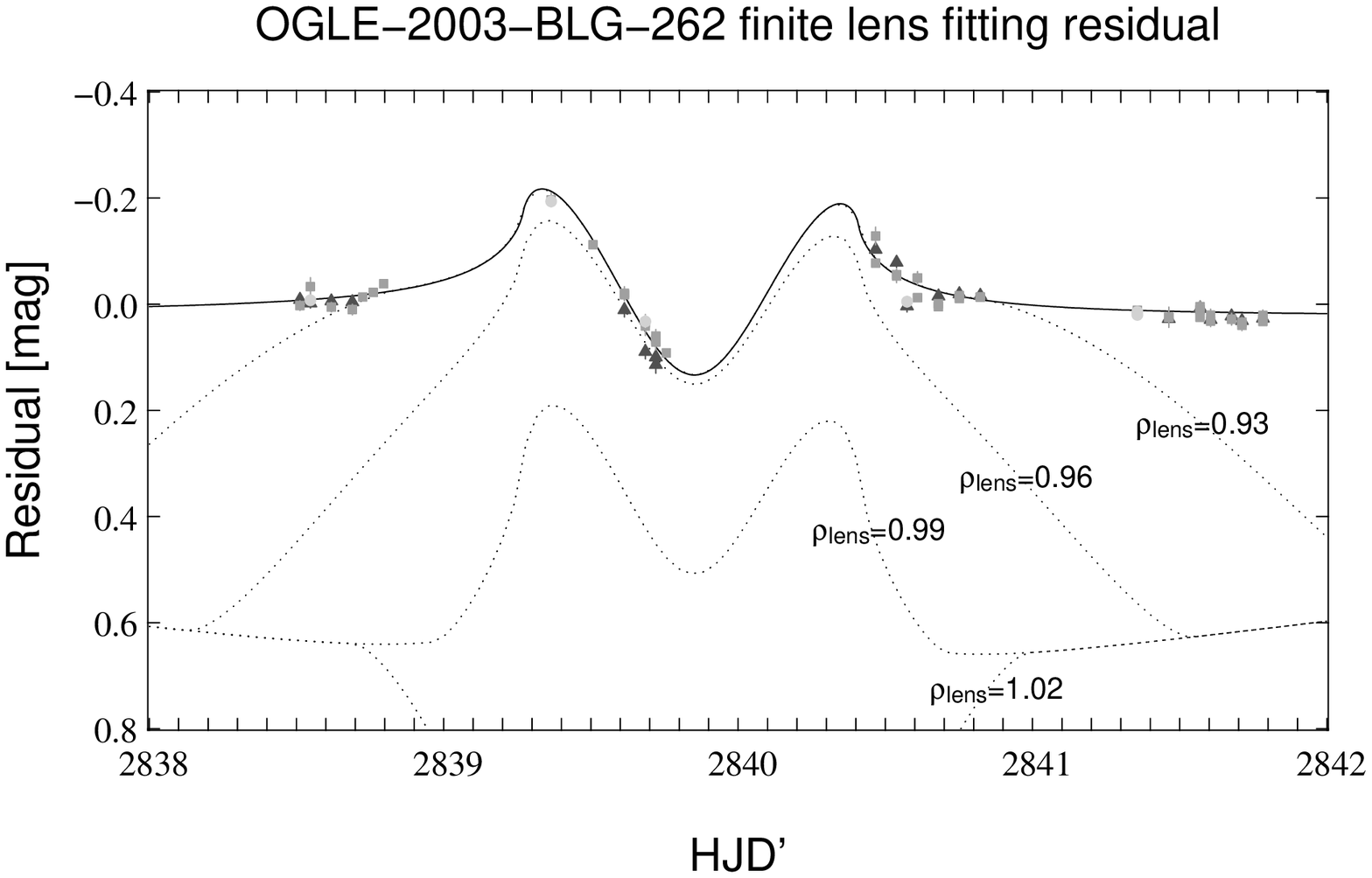}
  \caption{Fitting residual of various lens radius relative to the point-source 
model. Data points are in \textit{I}(square), \textit{V}(circle), and \textit{H}(triangle). 
The \textit{solid black line} shows the light curve for an extended source 
with uniform surface brightness. The \textit{dotted lines} illustrate the 
effects of finite lens sizes on top of finite source size for lens sizes of 
$\rholens$ = 0.93, 0.96, 0.99, and 1.02. One sees that all these cases can be safely excluded.
}
  \label{fig.OGLE262_RL_residual}
\end{minipage}
\end{figure}

\cite{1997ApJ...491..436A} then obtained the limb darkening coefficients of 
MACHO-1995-BLG-30 utilizing spectroscopic information. However, 
\cite{2003ApJ...594..464H} argued that the surface-brightness profile of this 
event can not be fully recovered due to its intrinsic complex variability. 
Therefore, we tested our limb-darkening fitting routine to another limb-darkened 
finite-source event OGLE-2003-BLG-262. Our results are shown in 
Table~\ref{tab.fit_OGLE262}, Figure~\ref{fig.OGLE262_residual} and Figure~\ref{fig.OGLE262} in comparison with 
\cite{2004ApJ...603..139Y}.

\begin{table}[!h]
\caption{Light-curve parameters for OGLE-2003-BLG-262.}
\footnotesize
\center
\begin{tabular}[t]{llllll}
\hline
Fit       & $A_{_{\mathrm{PS}}}(u)$                     & $\Afs(u;\RS)$                & $\Afs_{_{\mathrm{LD}}}(u;\RS)$                \\
\hline
t$_0$     & 2839.852 $\pm$ 0.001    & 2839.838 $\pm$ 0.001   & 2839.8361 $\pm$ 0.001 \\
t$_E$     & 12.83 $\pm$ 0.01        & 12.61 $\pm$ 0.01       & 12.559 $\pm$ 0.016   \\
u$_0$     & 0.02877 $\pm$ 0.00008   & 0.0365 $\pm$ 0.0002    & 0.0361 $\pm$ 0.0002    \\
$\RS$     & --                      & 0.0581 $\pm$ 0.0002    & 0.0598 $\pm$ 0.0002    \\
\hline
\multicolumn{4}{c}{\textit{Note.} We fixed the limb-darkening coefficients at $(\Gamma_V, \Gamma_I, \Gamma_H) = (0.72, 0.44, 0.26)$}
\end{tabular}
\label{tab.fit_OGLE262}
\end{table}
\normalsize

Finally, we choose several lens sizes for the configuration of OGLE-2003-BLG-262 to investigate the influence of the finite lens effects on 
the microlensing light curve in Figure~\ref{fig.OGLE262}.
The light curve is strongly altered only if the lens size is comparable to or larger than the Einstein radius.

\begin{figure}[!h]
  \centering
\epsscale{0.8}
\plotone{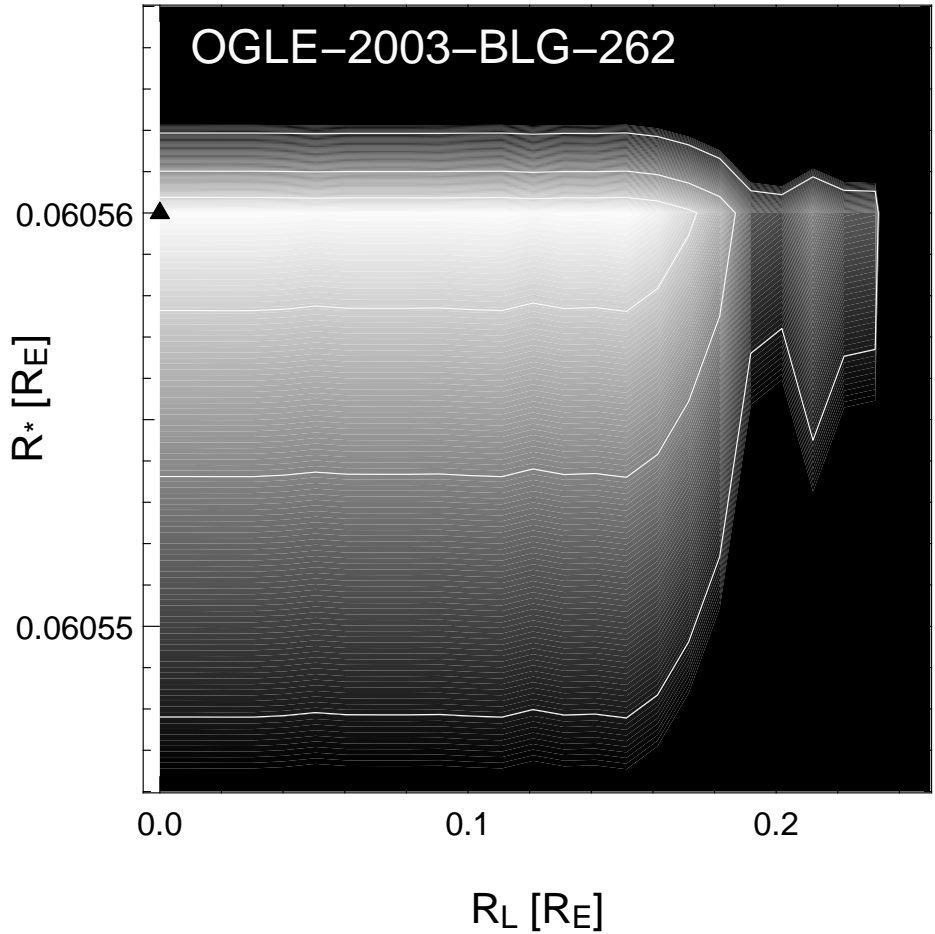}
  \caption{$\chi^2$ contour map of OGLE-2003-BLG-262. In \textit{white contour}, 
levels for 1, 2, and 3$\sigma$ for source and lens size fitting are shown. The \textit{black triangle} indicates the best-fitted model with $\RS$ = 0.06056 and $\Rlens$ = 0. This suggests that the point lens assumption is sufficient for OGLE-2003-BLG-262.}
  \label{fig.OGLE262_contour}
\end{figure}

When the lens size is smaller than 0.93$R_E$, it only partially covers the outer image and the finite lens 
effects can be observed only at the very beginning of the rising and near the end of the declining stage of the lensing event. 
Therefore, we fitted various lens sizes up to $\rholens = 1.1R_E$ using the full OGLE $I$-band data set. However, no improvement in $\chi^2$ has 
been found by introducing lens sizes as an extra parameter in the finite-source model (see Figure~\ref{fig.OGLE262_contour}). This implies 
that the lens size effect is negligible for OGLE-2003-BLG-262.  

\section{Conclusion}
We have demonstrated that finite-source effects can be more
conveniently evaluated in the lens-centered polar coordinate system.
The uniform source case can be reduced to a one-dimensional integral,
which can be solved in a fast and numerically stable manner. The
previously available formalisms were either comparably fast but held
only in the high-amplification regime (the Gould finite-source
approximation) or held in any amplification regime but involved an integral
which has singularity and is slower to solve (the Witt \& Mao approach).
We also showed that the vast majority of the OGLE-lensing events have
maximum amplifications smaller than 10, and therefore cannot be
precisely described in the high-amplification, finite-source approximation of Gould. 
Our formalism allows a fast and simultaneous search for microlensing
events with extended or pointlike sources in any amplification
regime.

We also presented the limb-darkening effects and finite lens size effects in
our formalism. We showed for the case of OGLE-2003-BLG-262 how one can constrain the
source size and obtain upper limits for the lens size.

The Appendix provides the partial derivatives of the amplification
for a uniform surface brightness source (Appendix A), a limb-darkened source (Appendix B), 
and a uniform surface brightness source with a finite lens (Appendix C), which are required 
in, e.g., the Levenberg-Marquardt algorithm to obtain microlensing light-curve fits.

\acknowledgements 
We thank Johannes Koppenh\"{o}fer for fruitful discussions. This work was supported by the DFG 
cluster of excellence `Origin and Structure of the Universe' (www.universe-cluster.de).

\clearpage
\appendix 
\section{A. Partial derivatives of the finite-source amplification for a source with uniform surface brightness}
\begin{equation}
  \begin{array}{ll}
  \frac{\partial\Afs}{\partial u} (u,\RS)= &
    \frac{2}{\pi \RS^2} \Int_0^\pi 
    \left\{
    \frac{u_2^2+2}{\sqrt{u_2^2+4}} 
    \left[\cos\vartheta-\frac{u\sin^2\vartheta}{\sqrt{\RS^2-u^2\sin^2\vartheta}} \right]
    -       
    \frac{u_1^2+2}{\sqrt{u_1^2+4}} 
    \left[\cos\vartheta+\frac{u\sin^2\vartheta}{\sqrt{\RS^2-u^2\sin^2\vartheta}} \right]
    \right\}
    d\vartheta 
\\

  \frac{\partial\Afs}{\partial \RS} (u,\RS)= &
    \frac{2}{\pi \RS} \Int_0^\pi  
    \left[
      \frac{u_2^2+2}{\sqrt{u_2^2+4}\sqrt{\RS^2-u^2\sin^2\vartheta}}
      -
      \frac{u_1^2+2}{\sqrt{u_1^2+4}\sqrt{\RS^2-u^2\sin^2\vartheta}}
      \right]
    \,d\vartheta 
    - 2\frac{\Afs(u,\RS)}{\RS}
~.
  \end{array}
  \label{eq.Afin_dRS}
\end{equation}

\section{B. Partial derivatives of the finite-source amplification for a source with limb darkening}

\begin{equation}
  \begin{array}{ll}
  \frac{\partial \Afs(u,\RS,\Gamma_{\lambda})}{\partial u}=&
    \frac{2}{\pi\RS^2}
    \Int_0^{\pi}
    \left[
      \frac{\partial u_2}{\partial u}
      \frac{u_2^2+2}{\sqrt{u_2^2+4}} 
      \frac{S_{\lambda}}{\bar{S}_{\lambda}}
      -
      \frac{\partial u_1}{\partial u}
      \frac{u_1^2+2}{\sqrt{u_1^2+4}} 
      \frac{S_{\lambda}}{\bar{S}_{\lambda}}
      \right]
    \,d\vartheta
    +
    \frac{2}{\pi\RS^2} 
    \Int_0^{\pi} \,
    \Int_{u_1}^{u_2} 
    \frac{\tilde{u}^2+2}{\sqrt{\tilde{u}^2+4}}\,
    \frac{\partial}{\partial u}
    \left(
    \frac{S_{\lambda}}{\bar{S}_{\lambda}}
    \right)\,d\tilde{u}d\vartheta
\\
  \frac{\partial \Afs(u,\RS,\Gamma_{\lambda})}{\partial \RS} =&
    \frac{2}{\pi\RS^2}
    \Int_0^{\pi}
    \left[
      \frac{\partial u_2}{\partial \RS}
      \frac{u_2^2+2}{\sqrt{u_2^2+4}} 
      \frac{S_{\lambda}}{\bar{S}_{\lambda}}
      -
      \frac{\partial u_1}{\partial \RS}
      \frac{u_1^2+2}{\sqrt{u_1^2+4}} 
      \frac{S_{\lambda}}{\bar{S}_{\lambda}}
      \right]
    \,d\vartheta
    + \frac{2}{\pi\RS^2}
    \Int_0^{\pi} \,
    \Int_{u_1}^{u_2}
    \frac{\tilde{u}^2+2}{\sqrt{\tilde{u}^2+4}}\,
    \frac{\partial}{\partial \RS}
    \left(
    \frac{S_{\lambda}}{\bar{S}_{\lambda}}
    \right)\,d\tilde{u}d\vartheta 
    -2\frac{\Afs(u,\RS,\Gamma_{\lambda})}{\RS}
\\
  \frac{\partial \Afs(u,\RS,\Gamma_{\lambda})}{\partial \Gamma_{\lambda}}= &
    \frac{2}{\pi\RS^2} 
    \Int_0^{\pi} \,
    \Int_{u_1}^{u_2}
    \frac{\tilde{u}^2+2}{\sqrt{\tilde{u}^2+4}}\,
    \frac{\partial}{\partial \Gamma_{\lambda}}
    \left(
    \frac{S_{\lambda}}{\bar{S}_{\lambda}}
    \right)\,d\tilde{u}
    \,d\vartheta
  \end{array}
  \label{eq.A_ld_dGamma}
\end{equation}

with 
$\frac{\partial u_1}{\partial u}=\cos\vartheta + u\sin^2\vartheta/\sqrt{\RS^2-u^2 \sin^2\vartheta}$,
$\frac{\partial u_2}{\partial u}=\cos\vartheta - u \sin^2\vartheta/\sqrt{\RS^2-u^2 \sin^2\vartheta}$,
$\frac{\partial u_1}{\partial \RS}= - \RS/\sqrt{\RS^2-u^2 \sin^2\vartheta}$,
$\frac{\partial u_2}{\partial \RS}= \RS/\sqrt{\RS^2-u^2 \sin^2\vartheta}$,
$\frac{\partial}{\partial u}\left(\frac{S_{\lambda}}{\bar{S}_{\lambda}}\right)=-\frac{3}{4} \Gamma_{\lambda}(-2\tilde{u}\cos\vartheta + 2u)/\left(\RS^2 \sqrt{1-\frac{\tilde{u}^2-2u\tilde{u}\cos\vartheta + u^2}{\RS^2}}\right)$,
$\frac{\partial}{\partial \RS}\left(\frac{S_{\lambda}}{\bar{S}_{\lambda}}\right)=\frac{3}{2} \Gamma_{\lambda}(\tilde{u}^2 - 2u\tilde{u}\cos\vartheta + u^2)/\left(\RS^3 \sqrt{1-\frac{\tilde{u}^2-2u\tilde{u}\cos\vartheta + u^2}{\RS^2}}\right)$,
$\frac{\partial}{\partial \Gamma_{\lambda}}\left(\frac{S_{\lambda}}{\bar{S}_{\lambda}}\right)=-1+\frac{3}{2} \sqrt{1-\frac{\tilde{u}^2-2u\tilde{u}\cos\vartheta+u^2}{\RS^2}}$ when $u_1$ and $u_2$ are not equal to zero.

\section{C. Partial derivatives of the finite-source and finite-lens amplification assuming a source with uniform brightness}
 \begin{equation}
   \begin{array}{rl}
 \frac{\partial\Afs}{\partial u}(u,\RS, \rholens) =
 &
 \scriptstyle{
   \frac{1}{\pi\RS^2}
   \Int_0^{\pi} \,
   \frac{\partial u_2}{\partial u}
   \left[
     \left(A_{_{\mathrm{PS}}}(u_2)+1\right)\Theta{\left(u_2-\rholens+\frac{1}{\rholens}\right)}
     +
     \left(A_{_{\mathrm{PS}}}(u_2)-1\right)\Theta{\left(-u_2-\rholens+\frac{1}{\rholens}\right)}
   \right] u_2
   \,d\vartheta
 }
       \\
 &
 \scriptstyle{
   -\frac{1}{\pi\RS^2}
   \Int_0^{\pi} \,
   \frac{\partial u_1}{\partial u}
   \left[
     \left(A_{_{\mathrm{PS}}}(u_1)+1\right)\Theta{\left(u_1-\rholens+\frac{1}{\rholens}\right)}
     +
     \left(A_{_{\mathrm{PS}}}(u_1)-1\right)\Theta{\left(-u_1-\rholens+\frac{1}{\rholens}\right)}
   \right] 
   u_1\,d\vartheta 
 }
 \\
 \frac{\partial\Afs}{\partial \RS}(u,\RS, \rholens) = 
 &
 \scriptstyle{ 
   \frac{1}{\pi\RS^2} 
   \Int_0^{\pi} \,
   \frac{\partial u_2}{\partial \RS}
   \left[
     \left(A_{_{\mathrm{PS}}}(u_2)+1\right)\Theta{\left(u_2-\rholens+\frac{1}{\rholens}\right)}
     +
     \left(A_{_{\mathrm{PS}}}(u_2)-1\right)\Theta{\left(-{u_2}-\rholens+\frac{1}{\rholens}\right)}
   \right] u_2\,d\vartheta 
 }
 \\
 &
 \scriptstyle{
   - 
   \frac{1}{\pi\RS^2}
   \Int_0^{\pi} \,
   \frac{\partial u_1}{\partial \RS}
   \left[
     \left(A_{_{\mathrm{PS}}}(u_1)+1\right)\Theta{\left(u_1-\rholens+\frac{1}{\rholens}\right)}
     +
     \left(A_{_{\mathrm{PS}}}(u_1)-1\right)\Theta{\left(-u_1-\rholens+\frac{1}{\rholens}\right)}
   \right] 
   u_1\,d\vartheta  
 }
 \\
 &
 \scriptstyle{
   -2\frac{\Afs(u,\RS, \rholens)}{\RS}
 }
 \\
 \frac{\partial\Afs}{\partial \rholens}(u,\RS, \rholens) =
 &
 \scriptstyle{
   \frac{1}{\pi\RS^2}
   \Int_0^{\pi} \,
   \Int_{u_1}^{u_2} 
   \left(A_{_{\mathrm{PS}}}(\tilde{u})+1\right)\delta{\left(\tilde{u}-\rholens+\frac{1}{\rholens}\right)} \frac{\partial \left(\tilde{u}-\rholens+\frac{1}{\rholens}\right)}{\partial \rholens}
   \tilde{u}\,d\tilde{u}\,d\vartheta    
 }
 \\
 &
 \scriptstyle{
   +  
   \frac{1}{\pi\RS^2}
   \Int_0^{\pi} \,
   \Int_{u_1}^{u_2} 
   \left(A_{_{\mathrm{PS}}}(\tilde{u})-1\right)\delta{\left(-\tilde{u}-\rholens+\frac{1}{\rholens}\right)} \frac{\partial \left(-\tilde{u}-\rholens+\frac{1}{\rholens}\right)}{\partial \rholens}
   \tilde{u}\,d\tilde{u}\,d\vartheta
 }
 ~.
 \end{array}
 \label{equ.dA-RL_dRL}
 \end{equation}

The derivatives can be obtained numerically by utilizing the same approache as 
shown in Equation~(\ref{eq.A_discret}). We also find that for $u>\RS$, substituting 
integration variable $\vartheta$ with $v\equiv\frac{u}{\RS}\sin{\vartheta}$ gives a 
numerically more stable estimations of the derivatives for a uniform brightness source:
\begin{equation}
  \begin{array}{ll}
  \frac{\partial\Afs}{\partial u} (u,\RS)= &
      \frac{u}{\pi\RS}   \Int_0^1 \frac{1}{U^3} \left[ 
        \frac{(u^2-\RS^2)(U+\Omega)-4\Omega} {\sqrt{(U+\Omega)^2+4}}  -
        \frac{(u^2-\RS^2)(U-\Omega)+4\Omega} {\sqrt{(U-\Omega)^2+4}}  
        \right] dv    
\\
  \frac{\partial\Afs}{\partial \RS} (u,\RS)= &
  \frac{1}{\pi\RS^2}   \Int_0^1 \frac{1}{U^3}\left[ 
    -\frac{  u^2\left( u^2-\RS^2 \right) (U+\Omega) - 4 (\Omega v^2\RS^2 - U^3)}
    {\sqrt{(U+\Omega)^2+4}}  +
    \frac{  u^2\left( u^2-\RS^2 \right) (U-\Omega) + 4 (\Omega v^2\RS^2 + U^3)}
         {\sqrt{(U-\Omega)^2+4}}   
         \right] dv 
  \end{array}
  \label{eq.Afin_dRS}
\end{equation}

with $U=\sqrt{u^2-v^2\RS^2}$ and $\Omega = \RS\sqrt{1-v^2}$.


\begin{thebibliography}{}
 \expandafter\ifx\csname natexlab\endcsname\relax\def\natexlab#1{#1}\fi
 
 \bibitem[{{Agol}(2002)}]{2002ApJ...579..430A}
 {Agol}, E. 2002: \textit{{Occultation and Microlensing}}, \apj, 579, 430
 
 \bibitem[Alcock et al.(1992)]{1992ASPC...34..193A} 
  {Alcock}, C., {Axelrod}, T.~S., {Bennett}, D.~P., {Cook}, K.~H., 
  {Park}, H.~S., {Griest}, K., {Perlmutter}, S., {Stubbs}, C.~W., 
  {Freeman}, K.~C., {Peterson}, B.~A. 1992: \textit{{The search for massive 
  compact halo objects with a (semi) robotic telescope}}, ASPC, 34, 193 

 \bibitem[{{Alcock} {\textit{et~al.}}(1997){Alcock}, {Allen}, {Allsman},
   {Alves}, {Axelrod}, {Banks}, {Beaulieu}, {Becker}, {Becker}, {Bennett},
   {Bond}, {Carter}, {Cook}, {Dodd}, {Freeman}, {Gregg}, {Griest}, {Hearnshaw},
   {Heller}, {Honda}, {Jugaku}, {Kabe}, {Kaspi}, {Kilmartin}, {Kitamura},
   {Kovo}, {Lehner}, {Love}, {Maoz}, {Marshall}, {Matsubara}, {Minniti},
   {Miyamoto}, {Morse}, {Muraki}, {Nakamura}, {Peterson}, {Phillips}, {Pratt},
   {Quinn}, {Reid}, {Reid}, {Reiss}, {Retter}, {Rodgers}, {Sargent}, {Sato},
   {Sekiguchi}, {Stetson}, {Stubbs}, {Sullivan}, {Sutherland}, {Tomaney},
   {Vandehei}, {Watase}, {Welch}, {Yanagisawa}, {Yoshizawa}, {Yock}, {The
   Macho}, \& {Gman Collaborations}}]{1997ApJ...491..436A}
   {Alcock}, C., {Allen}, W.~H., {Allsman}, R.~A., {Alves}, D., {Axelrod}, T.~S.,
   {Banks}, T.~S., {Beaulieu}, S.~F., {Becker}, A.~C., {Becker}, R.~H.,
   {Bennett}, D.~P., {Bond}, I.~A., {et~al.} 1997: \textit{{MACHO Alert 95-30:
   First Real-Time Observation of Extended Source Effects in Graviational
   Microlensing}}, \apj, 491, 436
 
 \bibitem[{{Cassan} {\textit{et~al.}}(2006){Cassan}, {Beaulieu}, {Fouqu{\'e}},
   {Brillant}, {Dominik}, {Greenhill}, {Heyrovsk{\'y}}, {Horne}, {J{\o}rgensen},
   {Kubas}, {Stempels}, {Vinter}, {Albrow}, {Bennett}, {Caldwell}, {Calitz},
   {Cook}, {Coutures}, {Dominis}, {Donatowicz}, {Hill}, {Hoffman}, {Kane},
   {Marquette}, {Martin}, {Meintjes}, {Menzies}, {Miller}, {Pollard}, {Sahu},
   {Wambsganss}, {Williams}, {Udalski}, {Szyma{\'n}ski}, {Kubiak},
   {Pietrzy{\'n}ski}, {Soszy{\'n}ski}, {{\.Z}ebru{\'n}}, {Szewczyk}, \&
   {Wyrzykowski}}]{2006A&A...460..277C}
   {Cassan}, A., {Beaulieu}, J.-P., {Fouqu{\'e}}, P., {Brillant}, S., {Dominik},
   M., {Greenhill}, J., {Heyrovsk{\'y}}, D., {Horne}, K., {J{\o}rgensen}, U.~G.,
   {Kubas}, D., {Stempels}, H.~C., {et~al.} 2006: \textit{{OGLE 2004-BLG-254: a
   K3 III Galactic bulge giant spatially resolved by a single microlens}}, \aap,
   460, 277
 
 \bibitem[{{Gould}(1994)}]{1994ApJ...421L..71G}
 {Gould}, A. 1994: \textit{{Proper motions of MACHOs}}, ApJ, 421, L71
 
 \bibitem[{{Gould}(1996)}]{1996ApJ...470..201G}
 {Gould}, A. 1996: \textit{Theory of Pixel Lensing}, ApJ, 470, 201+

\bibitem[{{Heyrovsky} \& {Loeb}(1997)}]{1997ApJ...490...38H}
  {Heyrovsky}, D. \& {Loeb}, A. 1997: \textit{{Microlensing of an Elliptical
      Source by a Point Mass}}, \apj, 490, 38

 \bibitem[{{Heyrovsk{\'y}}(2003)}]{2003ApJ...594..464H}
 {Heyrovsk{\'y}}, D. 2003: \textit{{Measuring Stellar Limb Darkening by
   Gravitational Microlensing}}, \apj, 594, 464
 
 \bibitem[{{Paczy{\'n}ski}(1986)}]{1986ApJ...304....1P}
 {Paczy{\'n}ski}, B. 1986: \textit{{Gravitational microlensing by the galactic
   halo}}, \apj, 304, 1
 
 \bibitem[{{Press} {\textit{et~al.}}(2007){Press}, {Teukolsky}, {Vetterling}, \& {Flannery}}]{2007nrc..book.....P}
 {Press}, W.~H., {Teukolsky}, S.~A., {Vetterling}, W.~T., {Flannery}, B.~P. 2007: Numerical recipes in C++ : the art of scientific
   computing (3rd ed.; Cambridge: Cambridge Univ. Press) 
 
 \bibitem[{{Riffeser} {\textit{et~al.}}(2006){Riffeser}, {Fliri}, {Seitz}, \&
   {Bender}}]{2006ApJS..163..225R}
 {Riffeser}, A., {Fliri}, J., {Seitz}, S., \& {Bender}, R. 2006:
   \textit{{Microlensing toward Crowded Fields: Theory and Applications to
   M31}}, ApJS, 163, 225
 
 \bibitem[Udalski et al.(1992)]{1992AcA....42..253U} 
  Udalski, A., Szymanski, M., Kaluzny, J., Kubiak, M., \& Mateo, M. 1992: \textit{{The Optical 
  Gravitational Lensing Experiment}}, AcA, 42, 253 

 \bibitem[{{Udalski}(2003)}]{2003AcA....53..291U}
 {Udalski}, A. 2003: \textit{{The Optical Gravitational Lensing Experiment. Real
   Time Data Analysis Systems in the OGLE-III Survey}}, Acta Astronomica, 53,
   291
 
 \bibitem[{{Witt} \& {Mao}(1994)}]{1994ApJ...430..505W}
 {Witt}, H.~J. \& {Mao}, S. 1994: \textit{{Can lensed stars be regarded as
   pointlike for microlensing by MACHOs?}}, \apj, 430, 505
 
 \bibitem[{{Yoo} {\textit{et~al.}}(2004){Yoo}, {DePoy}, {Gal-Yam}, {Gaudi},
   {Gould}, {Han}, {Lipkin}, {Maoz}, {Ofek}, {Park}, {Pogge}, {Udalski},
   {Soszy{\'n}ski}, {Wyrzykowski}, {Kubiak}, {Szyma{\'n}ski}, {Pietrzy{\'n}ski},
   {Szewczyk}, \& {{\.Z}ebru{\'n}}}]{2004ApJ...603..139Y}
 {Yoo}, J., {DePoy}, D.~L., {Gal-Yam}, A., {Gaudi}, B.~S., {Gould}, A., {Han},
   C., {Lipkin}, Y., {Maoz}, D., {Ofek}, E.~O., {Park}, B.-G., {Pogge}, R.~W.,
   {et~al.} 2004: \textit{{OGLE-2003-BLG-262: Finite-Source Effects from a
   Point-Mass Lens}}, \apj, 603, 139

\end{thebibliography}
\end{document}